\documentclass [11pt,a4paper] {article}

\usepackage[T1]{fontenc}
\usepackage{lmodern}

\usepackage[cp1252]{inputenc}
\usepackage{amssymb}
\usepackage{amsmath}
\usepackage{amsfonts,amssymb}
\usepackage[dvips]{graphicx}
\usepackage{bbm}
\usepackage{enumerate}

\usepackage[shortlabels]{enumitem}

\usepackage{amsthm}
\usepackage{cancel}

\usepackage{centernot}
\usepackage{mathrsfs}

\usepackage{wasysym}

\usepackage{xcolor}

\DeclareMathAlphabet{\mathpzc}{OT1}{pzc}{m}{it}

\begin{document}

\title{Finite geometry and black hole stability: Embedding discrete space into classical manifolds}

\author{Arkady Bolotin\footnote{$Email: arkadyv@bgu.ac.il$\vspace{5pt}} \\ \emph{Ben-Gurion University of the Negev, Beersheba (Israel)}}

\maketitle

\begin{abstract}\noindent The issue of defining the volume of black holes has significant implications for quantum gravity. Drawing on concepts from quantum theory and general relativity, several motivations for introducing discreteness in geometry can be proposed. However, to seriously consider any proposal for a discrete geometry, the identification problem and the challenge of defining a distance function within such a geometry must be addressed. This paper proposes the faithful embedding of sets representing spaces in finite geometry -- a specific type of discrete geometry characterized by a finite set of points -- into Riemannian manifolds as a solution to these problems. Similar to a classical measuring apparatus that interprets and understands quantum results in classical terms, classical geometry serves as a bridge between the discreteness of the physical world and our continuous understanding of the properties of space. In this framework, the volumetric density of information contained within a black hole is established, providing a consistent volume for the Schwarzschild black hole observed by all observers. Furthermore, the study finds that the minimum volume of the Schwarzschild black hole is non-zero. This fact implies that a black hole can only evaporate until its event horizon radius reaches the Planck length, signifying that black hole remnants are stable. Consequently, the total collapse of a black hole is prevented by the finite nature of the geometry describing physical space.\bigskip\bigskip

\bigskip\bigskip
\end{abstract}

\section{Introduction}  %{<-------------------------------------------------------------------------------------------------Section I}

\noindent The issue of defining the volume of black holes highlights a fundamental tension between our understanding of spacetime and the nature of black holes.\bigskip

\noindent In General Relativity (GR), the concept of volume is not as straightforward as in Newtonian physics because spacetime is curved and dynamic. The volume of a region in spacetime depends on how we slice it into spatial and temporal parts. This slicing can vary, leading to different definitions of volume for the same region \cite{Bini, Ruggiero, Kraniotis}. Consequently, within the framework of GR, discussing a spatial volume appears to lack meaningful context.\bigskip

\noindent However, this state of the matter is unsatisfactory for a black hole.\bigskip

\noindent First, this is because the area of the event horizon of a black hole is an invariant quantity. But if the area of the boundary of a black hole remains the same for all observers, how can the volume of the black hole be non-invariant?\bigskip

\noindent Second, it is due to the Bekenstein-Hawking (BH) entropy bound. The formulation of the BH entropy bound starts with a choice of spatial volume $V$. The volume, in turn, defines a boundary $\delta V$, whose area $A$ is then claimed to be an upper bound on $H(V)$, the entropy in $V$ \cite{Bousso}. Specifically,\smallskip

\begin{equation} \label{BHB} %{Eq.1}
  H\!\left( V\right)
   \le
   \frac{A\left( \delta V\right)}{4 \ell_P^2}
   \;\;\;\;   
\end{equation}
\smallskip

\noindent where $\ell_P$ stands for the Planck length. This begs the question: If $V$ were to be meaningless, how could the BH entropy bound hold true?\bigskip

\noindent Several approaches have been proposed to define a meaningful volume for black holes. One idea is to postulate that spacetime admits the existence of a timelike Killing vector in some regions (for instance, in the Schwarzschild solution, which describes the spacetime outside a spherical, non-rotating mass, there is a timelike Killing vector corresponding to the time translation symmetry of the static black hole). This would imply that it is possible to define a meaningful notion of volume, even in the absence of a globally timelike Killing vector \cite{Parikh}.\bigskip

\noindent Another approach involves rethinking the concept of volume inside a black hole. Unlike the static exterior described by the Schwarzschild metric, the interior can have a time-dependent notion of volume. For a collapsed object, this volume would grow with time since the collapse, reaching a simple asymptotic form that has a compelling geometrical interpretation \cite{Rovelli}.\bigskip

\noindent However, those approaches face challenges in aligning with the intuition behind the BH entropy bound. For instance, when defining the volume of black holes using a somewhere-timelike Killing vector, one might ask: Does the entropy account for all possible gravitational degrees of freedom within the volume $V$ enclosed by the boundary $\delta V$ of area $A(\delta V)$? If it does, then this implies that quantum gravity is highly nonlocal, possessing far fewer degrees of freedom than a local quantum field theory would suggest. Conversely, if it does not, then quantum gravity is local, implying that the field deep inside a black hole is not entangled with the field outside.\bigskip

\noindent Additionally, in the approach that defines the volume of black holes as time-dependent, a different problem arises: as the interior volume of a black hole $V$ grows over time, what implications does this have for the entropy contained within  $V$?\bigskip

\noindent These problems indicate that the relationship between entropy and the geometry of spacetime is not fully understood.\bigskip

\noindent As a result, the issue of defining the volume of black holes has significant implications for quantum gravity. In the geometric theory of gravitation, black holes are described by their event horizons, and properties like entropy are related to the area of these horizons. However, in quantum field theory (QFT), entropy increases with volume. The challenge, therefore, is to reconcile the extensive nature of entropy in QFT with the area law of entropy in GR. Achieving such a reconciliation could help to extend a quantum field theoretic description to curved spacetimes, such as around black holes or in cosmological models.\bigskip

\section{Materials and Methods}

\subsection{Problems in Addressing the Discrete Structure of Spacetime}  %{<------------------------------------------------------------------------------------------------- I}

\noindent There are various compelling reasons for advocating discreteness and finitism in geometric theories.\bigskip

\noindent One motivation is derived from quantum theory, which introduces the hypothesis of quantization -- the notion that the magnitude of a physical property can assume only discrete values. This hypothesis suggests that space itself may have a discrete structure \cite{Rovelli}.\bigskip

\noindent Another compelling motivation is that the discreteness of space could act as a natural ultraviolet cutoff, helping to regularize divergences in physical theories \cite{Reitz}.\bigskip

\noindent Moreover, consistent with the black hole entropy bound (\ref{BHB}), only a finite number of degrees of freedom can be localized within any finite region of space \cite{Singh, Bolotin}. Since, in QFT, degrees of freedom are inherently tied to the fundamental elements of space -- namely, points -- this implies that the number of points within any finite-volume region must necessarily be finite.\bigskip

\noindent However, to take any proposal for discrete geometry seriously, \emph{the problem of defining a distance function} within such a geometry needs to be solved. As a matter of fact, there is a rather devastating argument that shows the impossibility of a genuine distance function for discrete geometry. This argument, which was first formulated by Hermann Weyl \cite{Weyl}, is surprisingly simple and can be summarized as follows: If a square is constructed from fundamental tiles (e.g., with each tile having an area of the square Planck length, $\ell_P^2$), then the number of tiles along the diagonal is the same as the number along the sides; thus, the diagonal would be equal in length to the side.\bigskip

\noindent Although at least three solutions to this problem have been proposed (see details, for example, in \cite{Bendegem}), they all indicate that without classical geometry in the background, it is impossible to construct a genuine distance function. Besides, these solutions do not address the question of how classical geometry can re-emerge, even approximately.\bigskip

\noindent Another troubling complication with discrete geometry is \emph{the identification problem}. To understand it, recall that unlike in continuous geometry (such as classical Euclidean geometry), where a point has no size, area, volume, or any other measurable attribute, a point in discrete geometry must possess one or more measurable properties. Hence, the question arises: Which measurable attribute(s) should be identified with points in this geometry?\bigskip

\noindent This paper will demonstrate a way to resolve both problems associated with discrete geometry.\bigskip

\subsection{Embedding Finite Geometries into Classical Frameworks}  %{<------------------------------------------------------------------------------------------------- II}

\noindent We start by proposing that physical space is fundamentally discrete and can be modeled using finite geometry -- a specialized form of discrete geometry characterized by a finite set of points. Furthermore, each point in this set is treated as a unit of information, resonating with Wheeler's concept of `It from Bit' \cite{Wheeler}.\bigskip

\noindent Another argument in favor of identifying points of finite geometry as informational entities is the inherently discrete nature of information -- unlike other physical properties. Information is typically quantified in bits, which are fundamentally discrete units. Consequently, linking points to information aligns naturally with the intrinsic discreteness of finite geometry.\bigskip

\noindent Let $\{M\}$ denote the set of points representing a space $M$ in this geometry (i.e., a space that is characterized by a discrete and finite structure, as opposed to the continuous and infinite nature of classical geometry). The cardinality $|\{M\}|$ (the number of points in $M$) will then be equal to the amount of information contained in $M$, which, in other words, is the entropy in $M$:\smallskip

\begin{equation}  %{Eq.2}
  H\!\left( M\right)
   \equiv
   k_B
   \cdot
   \left|\{M\}\right|
   \;\;\;\;   
\end{equation}
\smallskip

\noindent where $k_B$ stands for the Boltzmann constant.\bigskip

\noindent Consider a Riemannian manifold $\mathfrak{R}$. This manifold is a type of geometric space, i.e., a mathematical structure used in the theoretical and axiomatized studies of physical space, e.g., in the geometric theory of gravitation such as GR. The manifold $\mathfrak{R}$ is smooth and equipped with a Riemannian metric, which allows for the measurement of distances and angles. This metric provides a way to define various geometric concepts such as lengths of curves, angles between vectors, and volumes of regions. Let $R_{\mathfrak{R}}$ be a region of $\mathfrak{R}$ and $\delta R_{\mathfrak{R}}$ its boundary, for which measures like the volume $V(R_{\mathfrak{R}})$ and the area $A(\delta R_{\mathfrak{R}})$ are defined, respectively.\bigskip

\noindent Assume that the set $\{M\}$ is \emph{faithfully embedded} into the manifold $\mathfrak{R}$. This implies that the number of points in $\{M\}$ mapped into the region $R_{\mathfrak{R}}$ and its boundary $\delta R_{\mathfrak{R}}$ is proportional to the corresponding measure -- namely, the volume $V(R_{\mathfrak{R}})$ or the area $A(\delta R_{\mathfrak{R}})$. Since points of $\{M\}$ are identified with bits of information, the entropies within $V(R_{\mathfrak{R}})$ and $A(\delta R_{\mathfrak{R}})$ can be expressed as follows:\smallskip

\begin{equation}  %{Eq.3}
   H\!\left( V\left(R_{\mathfrak{R}}\right)\right)
   =
   k_B
   \cdot
   \rho_V
   \cdot
   V\left(R_{\mathfrak{R}}\right)
   \;\;\;\;   
\end{equation}
\vspace{-20pt}

\begin{equation} \label{EA} %{Eq.4}
   H\!\left( A\left(\delta R_{\mathfrak{R}}\right)\right)
   =
   k_B
   \cdot
   \rho_A
   \cdot
   A\left(\delta R_{\mathfrak{R}}\right)
   \;\;\;\;   
\end{equation}
\smallskip

\noindent Here, $\rho_V$ and $\rho_A$ denote the volumetric and areal density of information contained within the region $R_{\mathfrak{R}}$ and its boundary $\delta R_{\mathfrak{R}}$, respectively.\bigskip

\noindent Faithful embedding can also be conceptualized as randomly sprinkling bits of information -- analogous to dropping ink droplets -- using a Poisson process within a Riemannian manifold $\mathfrak{R}$. If information is sprinkled into a region $R_{\mathfrak{R}}$ and its boundary $\delta R_{\mathfrak{R}}$ in proportion to the volume $V(R_{\mathfrak{R}})$ and the area $A(\delta R_{\mathfrak{R}})$, the parameters $\rho_V$ and $\rho_A$ can be regarded as the volumetric and areal densities of the sprinkling, respectively.\bigskip

\noindent It is intriguing to observe that the challenge of defining a distance function within finite geometry closely parallels the measurement problem in quantum mechanics. In quantum mechanics, a classical measuring apparatus is indispensable for interpreting and translating quantum results into classical terms, enabling the connection between quantum phenomena and our everyday classical experiences. Likewise, classical geometry functions as a crucial intermediary between the discrete nature of the physical world and our persistent comprehension of the properties of space. Without classical geometry, bridging discrete phenomena with the continuous domain of our everyday experience would be immensely difficult.\bigskip

\subsection{Discreteness and the Einstein Equivalence Principle: Implications for Black Hole Physics}  %{<------------------------------------------------------------------------------------ III}

\noindent If $R_{\mathfrak{R}}$ is the region of a black hole and $\delta R_{\mathfrak{R}}$ is the event horizon (the boundary of the region of a black hole), then the amount of entropy assigned to the black hole to comply with the laws of thermodynamics as interpreted by external observers (known as the Bekenstein-Hawking entropy or black hole entropy \cite{Bekenstein}, \cite{Hawking}) is given by\smallskip

\begin{equation} \label{BH} %{Eq.5}
   S_{BH}
   =
   k_B
   \cdot
   \frac{1}{4 \ell_P^2}
   \cdot
   A\left(\delta R_{\mathfrak{R}}\right)
   \;\;\;\;   
\end{equation}
\smallskip

\noindent where $ A(\delta R_{\mathfrak{R}})$ denotes the area of the event horizon.\bigskip

\noindent Assuming that the black hole entropy $S_{BH}$ coincides (up to Boltzmann's constant) with the informational content of the event horizon $\delta R_{\mathfrak{R}}$, we find that $S_{BH}$ aligns with $H(A(\delta R_{\mathfrak{R}}))$, the entropy in $\delta R_{\mathfrak{R}}$.\bigskip

\noindent Comparing Eq. (\ref{BH}) with Eq. (\ref{EA}) immediately yields the formula for the areal density of information:\smallskip

\begin{equation}  %{Eq.6}
   \rho_A
   =
   \frac{1}{4 \ell_P^2}
   \;\;\;\;   
\end{equation}
\smallskip

\noindent As to the volumetric density of information $\rho_V$, in order for the BH entropy bound (\ref{BHB}) to hold true in faithful embedding, the density $\rho_V$ must meet the condition\smallskip

\begin{equation}  \label{COND} %{Eq.7}
   \rho_V
   \cdot
   V\!\left(R_{\mathfrak{R}}\right)
   \le
   \rho_A
   \cdot
   A\left(\delta R_{\mathfrak{R}}\right)
   \;\;\;\;   
\end{equation}
\smallskip

\noindent This can only be achieved if\smallskip

\begin{equation}  %{Eq.8}
   \rho_V
   \le
   \frac{1}{4 \ell_P^2}
   \cdot
  \frac{ A\left(\delta R_{\mathfrak{R}}\right)}{V\!\left(R_{\mathfrak{R}}\right)}
   \;\;\;\;   
\end{equation}
\smallskip

\noindent As acknowledged, on large scales, the geometric space attributed to the universe is well approximated as three-dimensional and flat \cite{Baumann}. Given that, the Riemannian manifold $\mathfrak{U}$ associated with the universe can be considered to have a flat geometry, meaning its Riemannian metric corresponds to the standard Euclidean space. Similarly, the region $R_{\mathfrak{U}}$ of $\mathfrak{U}$ representing the geometric space of the observable universe can be viewed as a 3-dimensional Euclidean ball with the radius $r_{\mathfrak{U}}$.\bigskip

\noindent Provided that the ratio of the area of the boundary of $R_{\mathfrak{U}}$ to the volume of $R_{\mathfrak{U}}$ is\smallskip

\begin{equation}  %{Eq.9}
   \frac{ A\left(\delta R_{\mathfrak{U}}\right)}{V\!\left(R_{\mathfrak{U}}\right)}
   =
   \frac{3}{r_{\mathfrak{U}}}
   \approx
   \frac{H_0}{c}
   \;\;\;\;   
\end{equation}
\smallskip

\noindent where $H_0$ represents the Hubble constant (i.e., the present value of the Hubble parameter $H$) \cite{Peebles}, the informational density $\rho_V$ has an upper limit. Specifically,\smallskip

\begin{equation} \label{LIM} %{Eq.10}
   \max{\rho_V}
   \approx
   \frac{1}{4 \ell_P^2}
   \cdot
   \frac{H_0}{c}   
   \;\;\;\;   
\end{equation}
\smallskip

\noindent Since this limit is expressed solely in terms of physical constants, it remains invariant for all observers. Furthermore, it can be argued that this limit is also preserved within the geometric space inside a black hole.\bigskip

\noindent This invariance arises because, for bits of information to be faithfully embedded within any subregion of $R_{\mathfrak{U}}$, the informational density $\rho_V$ must be independent of the specific subregion. In other words, $\rho_V$ must be location-independent. When this condition is satisfied, the distribution of information within the region $R_{\mathfrak{U}}$ will be uniform, adhering to the isotropy of the region.\bigskip

\noindent More specifically, the probability that a subregion $R \subset R_{\mathfrak{U}}$ contains $n$ bits of information can be described by the homogeneous (uniform) Poisson point process:\smallskip

\begin{equation} \label{POS} %{Eq.11}
   P\left( n \right)
   =
   \frac{\left( \rho_V \cdot V(R)\right)^n\cdot e^{-\rho_V \cdot V(R)}}{n!}
   \;\;\;\;   
\end{equation}
\smallskip

\noindent Here, the parameter $\lambda = \rho_V \cdot V(R)$ denotes the expected number of bits assigned to the subregion $R$. The homogeneous Poisson point process is invariant to the position within the underlying geometric space via its parameter $\lambda$, making it both a stationary process (invariant under translation) and an isotropic process (invariant under rotation).\bigskip

\noindent When discussing isotropy of a geometric space inside a black hole, it is important to note that while spacetime near the event horizon is highly curved, the curvature remains symmetric around the black hole's center. Thus, as long as one does not approach the singularity -- the central point of infinite density in classical black hole models -- isotropy will generally remain intact.\bigskip

\noindent An exception to this arises in the case of Kerr black holes (rotating black holes), where rotation introduces an axis of symmetry, thereby disrupting isotropy.\bigskip

\noindent Thus, disregarding the influence of the singularity or rotating black holes, the informational density $\rho_V$ can reasonably be considered location-independent. As a result, the distribution of information within a black hole may be assumed to reflect that outside, including the upper limit (\ref{LIM}).\bigskip

\section{Results}

\subsection{From Finite Geometry to Stable Black Hole Remnants}  %{<--------------------------------------------------------------------------------------- I}

\noindent We are now ready to define the volume of a black hole.\bigskip

\noindent For simplicity, assume that the event horizon of the black hole is spherical, which holds true for a Schwarzschild black hole. Applying the Eq. (\ref{COND}) and (\ref{LIM}), the upper limit of the volume of the Schwarzschild black hole can be expressed as:\smallskip

\begin{equation} \label{VBH} %{Eq.12}
   \max{V\!\left(R_{\mathfrak{R}}\right)}
   \approx
   4 \pi \ell_H r_{S}^2
   \;\;\;\;   
\end{equation}
\smallskip

\noindent where $\ell_H = cH_{0}^{-1}$ represents the Hubble length and $r_{S}$ is the Schwarzschild radius. Since all the terms in the above expression are invariant quantities, the calculated upper limit, $\max{V\!\left(R_{\mathfrak{R}}\right)}$, is the same for all observers.\bigskip

\noindent Given that the Hubble length $\ell_H$ is approximately $1.32 \times 10^{26}$ meters \cite{Planck}, this upper bound is extraordinarily large. For example, for a black hole with a Schwarzschild radius $r_{S}$ of just 3 km, $\max{V\!\left(R_{\mathfrak{R}}\right)}$ corresponds to the volume of a regular sphere $\frac{4}{3} \pi r_{\text{sphere}}^3$ with the radius:\smallskip

\begin{equation}  %{Eq.13}
   r_{\text{sphere}}
   =
   \sqrt[3] {3\ell_H^2 r_{S}^2}
   \approx
   3.32 \times 10^{10} \, \text{meters}
   \;\;\;\;   
\end{equation}
\smallskip

\noindent which exceeds the distance from the Sun to Mars, approximately 225 million kilometers.\bigskip

\noindent Let us find the minimum volume of a Schwarzschild black hole, $V_{\text{min}}(R_{\mathfrak{R}})$. Assuming that the parameter $\lambda$ of the Poisson distribution (\ref{POS}) is given by\smallskip

\begin{equation}  %{Eq.14}
   \lambda
   =
   \max{\rho_V}
   \cdot
   V_{\text{min}}\!\left(R_{\mathfrak{R}}\right)
   \;\;\;\;   
\end{equation}
\smallskip

\noindent it follows that as $\lambda$ approaches zero, the probability of having zero information within the volume $V_{\text{min}}(R_{\mathfrak{R}})$ approaches 1, while the probability of having any non-zero number of bits of information approaches 0. This indicates that when the volume $V_{\text{min}}(R_{\mathfrak{R}})$ is very small, it becomes nearly certain that no entropy can be assigned to it.\bigskip

\noindent However, this scenario is unacceptable, as the absence of entropy implies zero degrees of freedom, effectively eliminating quantum fluctuations. Thus, it is reasonable to propose that the volume $V_{\text{min}}(R_{\mathfrak{R}})$ must be such that the expected number of bits of information it contains is equal to 1. This leads to the definition of $V_{\text{min}}(R_{\mathfrak{R}})$ as:\smallskip

\begin{equation}  %{Eq.15}
   V_{\text{min}}\left(R_{\mathfrak{R}}\right)
   =
   \left(\max{\rho_V} \right)^{-1}
   \approx
   4 \ell_H \ell_{P}^2
   \;\;\;\;   
\end{equation}
\smallskip

\noindent This volume corresponds to that of a regular sphere with a radius $r_{\text{sphere}} \approx 1.49 \times 10^{-15}$ meters, which falls within the typical range of the proton/neutron radius.\bigskip

\noindent Comparing $V_{\text{min}}(R_{\mathfrak{R}})$ with Eq. (\ref{VBH}) suggests that a Schwarzschild black hole with the smallest possible interior volume would have a Schwarzschild radius on the order of the Planck length. Specifically, $r_{S} =\frac{\ell_{P}}{\sqrt{\pi}} \approx 0.9 \times 10^{-35}$ meters.\bigskip

\noindent The fact that the minimum volume $V_{\text{min}}(R_{\mathfrak{R}})$ is nonzero indicates that a Schwarzschild black hole can only evaporate until its event horizon radius reduces to the Planck length. As $V_{\text{min}}(R_{\mathfrak{R}})$ is defined solely in terms of three physical constants -- $c$, $\ell_P$, and $H_0$ -- it follows that black hole evaporation remnants are inherently stable. Therefore, the stability of black hole remnants (BHRs) finds its explanation in the finite nature of the geometry that underpins physical space.\bigskip

\subsection{Discussion}  %{<--------------------------------------------------------------------------------------- II}

\noindent In the literature of the last 50 years, two unresolved issues of black hole physics are actively discussed: the singularity and the final state of black hole evaporation, specifically, BHRs \cite{Bonanno}. Regarding the latter, BHRs are theorized to be stable due to a combination of quantum mechanical and gravitational principles \cite{Belfaqih}. Key reasons for this include considerations of the minimal size of a black hole and the presence of quantum effects \cite{Wang}.\bigskip

\noindent In more detail, as the black hole evaporates through Hawking radiation, it shrinks. When it reaches the size comparable to the Planck length, further evaporation would require a violation of the principles of quantum mechanics and GR, suggesting a lower limit to how small the black hole can get. At the Planck scale, quantum gravitational effects become significant, preventing the black hole from shrinking further. These effects could stabilize the remnant by halting further evaporation \cite{Russ}.\bigskip

\noindent The existence of BHRs is also deduced from the Generalized Uncertainty Principle (GUP), an extension of the Heisenberg Uncertainty Principle that incorporates gravitational effects to refine the limits of measurement precision within quantum mechanics \cite{Bosso, Tee, Escors, Casadio, Nozari}. GUP, in turn, is inferred from the study of string collisions at Planckian energies in string theory.\bigskip

\noindent However, it has been shown in Ref. \cite{Lobos, Bang} that GUP can be derived without considering strings, but rather through a thought experiment measuring the radius of the apparent horizon of a black hole.\bigskip

\noindent Similarly, the present study demonstrates that the stability of BHRs can be understood without invoking advanced theories of quantum gravity, such as string theory or loop quantum gravity. Instead, it uses very general and model-independent considerations based on the finite geometrical structure of physical space. This approach is advantageous as it can presumably be fulfilled by any candidate quantum theory of gravitation.\bigskip

\noindent It is worth noting that if a sufficient number of small black holes were formed in the early universe, the resulting BHRs could serve as a compelling candidate for dark matter. Direct observation of BHRs, however, appears unlikely due to their purely gravitational mode of interaction. Nevertheless, as black hole evaporation halts upon reaching a BHR, the spectrum of gravitons emitted during this process should exhibit a cutoff \cite{Chen05}. This cutoff could potentially offer an observational signature of discrete space linked to the cosmic gravitational wave background.\bigskip

\section*{Funding}

\noindent The author declares no financial support for the research, authorship, or publication of this article.\bigskip

\section*{Author contributions}

\noindent A.B. is solely responsible for the conception, research, writing, and finalization of this article.\bigskip

\section*{Data Availability Statement}

\noindent The original contributions presented in the study are included in the article, further inquiries can be directed to the corresponding author.\bigskip

\section*{Conflict of interest}

\noindent The author states that there is no conflict of interest.\bigskip

\section*{Institutional review board statement}

\noindent Not applicable.\bigskip

\section*{Informed consent statement}

\noindent Not applicable.\bigskip

\bibliographystyle{References}

\end{document}